\documentstyle[newarcrc,fleqn,epsfig]{article}

\title{{\it HST} Observations of GRO J1655--40 in Outburst}

\author{Robert I. Hynes\address{Astronomy Centre, University of
	Sussex, Falmer, Brighton, East Sussex, BN1 9QJ, United Kingdom}}

\begin{document}
\maketitle

\begin{abstract}

We examine the results of a coordinated {\it HST}--{\it RXTE}--{\it
CGRO} campaign to study the microquasar GRO J1655--40 during its
1996--7 outburst, focusing on interpretation of the overall
anti-correlation seen between optical and X-ray fluxes during the
early months of the outburst.  Our tools include echo-mapping,
optical/UV continuum spectral modelling and analysis of spectral
variability.  We conclude by suggesting one possible interpretation
for the anti-correlation.

\end{abstract}
\section{Introduction}
\begin{figure}[htb]
\begin{minipage}[t]{2.8in}
\epsfig{width=2.5in,file=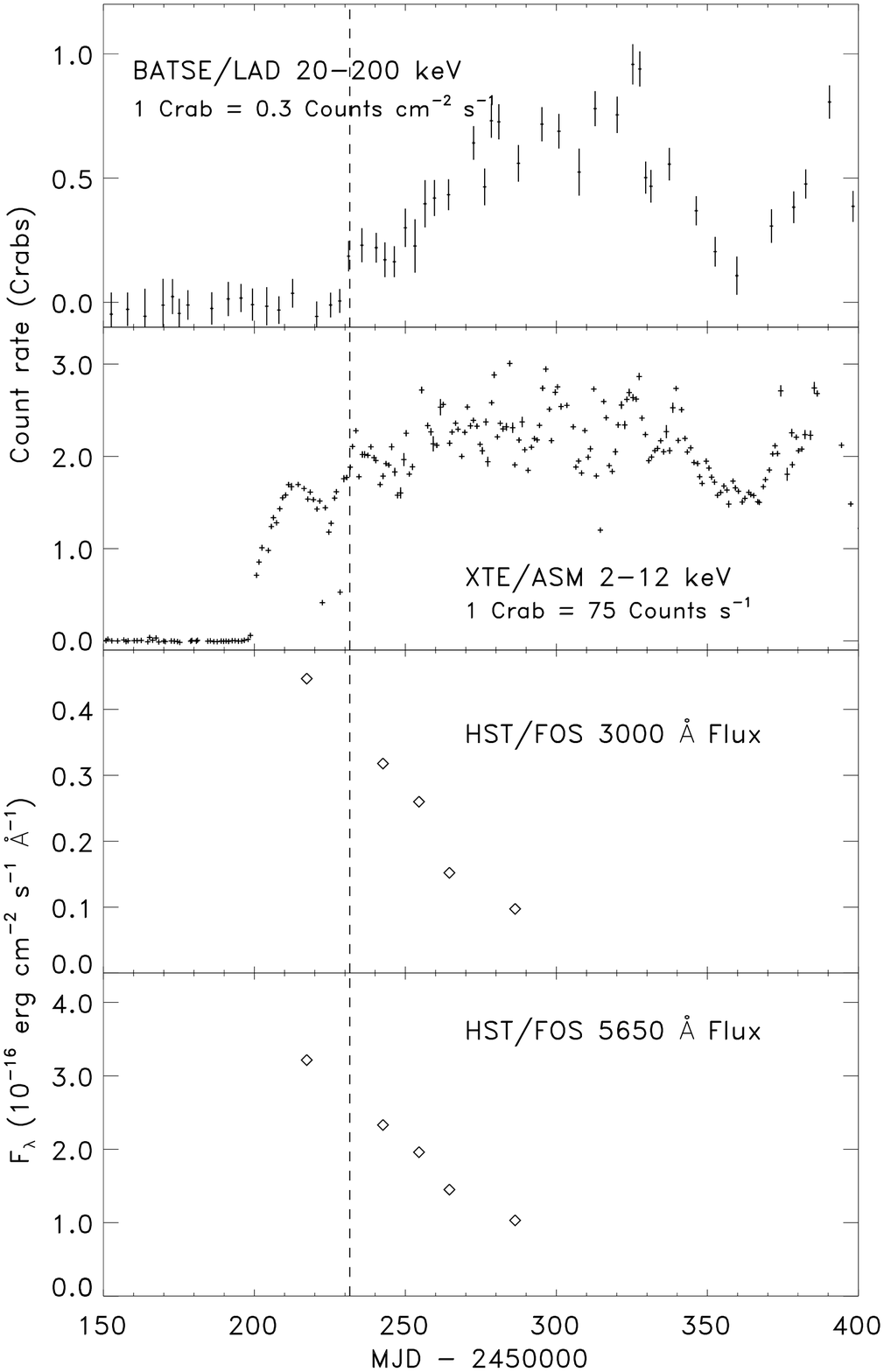}
\label{LongLCFig}
\end{minipage}
\hspace{\fill}
\begin{minipage}[b]{3.3in}
\caption{ (left) Light curves of first part the 1996--7 outburst.  The
         time-axis begins at 1996 March 8.  The discrepant behaviour
         of the optical-UV and X-ray data is clear.  The dotted line
         shows the first radio detection of this outburst (Hunstead
         and Campbell-Wilson 1996.)}
\vspace*{0.15in}
\epsfig{width=2.4in,angle=90,file=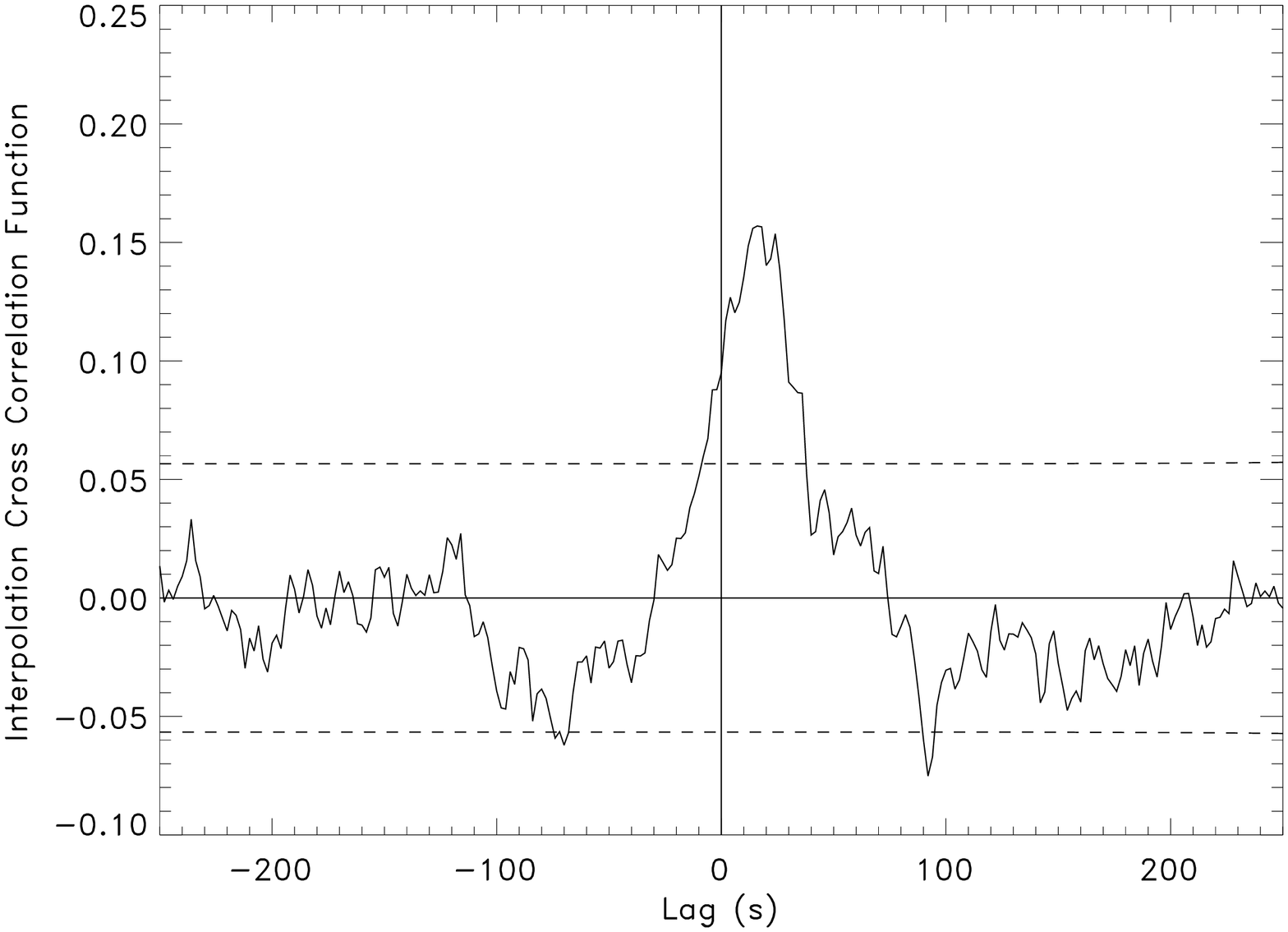}
\vspace*{-0.5in}
\caption{The combined interpolation cross-correlation function of our
         June 8 {\it HST/RXTE} data.  Dashed lines denote $3\sigma$
         limits on spurious correlations.}
\label{ICFFig}
\end{minipage}
\end{figure}

During the early stages of the 1996--7 outburst of the microquasar GRO
J1655--40, {\it HST}, {\it RXTE} and {\it CGRO} light curves reveal an
enigmatic behaviour, shown in Fig.\ \ref{LongLCFig}.  Over a period of
about 3 months, during which optical and UV fluxes declined steadily,
the X-ray brightness of the object {\em increased}.  The hard X-ray
rise in particular appears almost {\em anti-correlated} with the
optical/UV decline; this is difficult to reconcile with models in
which the optical/UV flux is produced by reprocessing in an irradiated
accretion disc, e.g.\ King \& Ritter (1998), for which the optical
flux is expected to track X-ray behaviour.  Nonetheless, comparison of
optical and X-ray light curves reveals correlated variability; echo
mapping suggests that this is due to the disc being significantly
irradiated, contrary to what the long term light curves would suggest.

Our {\it HST} data set spans 1996 May 14 to July 22, comprising five
separate observations, during four of which the {\it RXTE}/PCA also
observed the source.  Each observation resulted in a series of spectra
(spanning most or all of the 1300--9000\,\AA\ range) obtained with time
resolution of 2--3\,s.  We will examine the evidence which these data
provide for reprocessed X-rays being an important source of optical
flux and suggest one way to reconcile this evidence with the
apparently contradictory longer term behaviour.  The hard X-ray vs.\
optical/UV anti-correlation may arise naturally from this interpretation.

\section{RAPID Spectroscopy -- an Exercise in Echo Mapping}
Our {\it HST/RXTE} data from June 8 (photometric phase 0.4) shows
correlated short-term variability (on timescales of seconds to
minutes), with the X-ray variations leading the optical by 10--20
seconds.  The combined interpolation cross correlation function is
shown in Fig.\ \ref{ICFFig}.  We interpret this correlation as due to
reprocessing of a variable X-ray flux into optical/UV emission and
hence perform echo mapping of the reprocessing region (Hynes et al.\
1998a.)  The observed lags are too short to be consistent with
reprocessing on the companion star (at this orbital phase causality
requires a minimum lag of $\sim 40$\,s from the companion) but are of
the size expected for echoes from the accretion disc.  We therefore
conclude that X-rays are being reprocessed into optical/UV photons in
the disc.  We estimate that to produce the amplitude of variability
observed by {\it HST}, at least 15--20\,\% of the optical flux must be
generated in this way.
\section{Rapid SPECTROSCOPY -- Making Sense of Continuum Spectra}
\begin{figure}[htb]
\begin{minipage}[t]{3.0in}
\epsfig{width=2.1in,angle=90,file=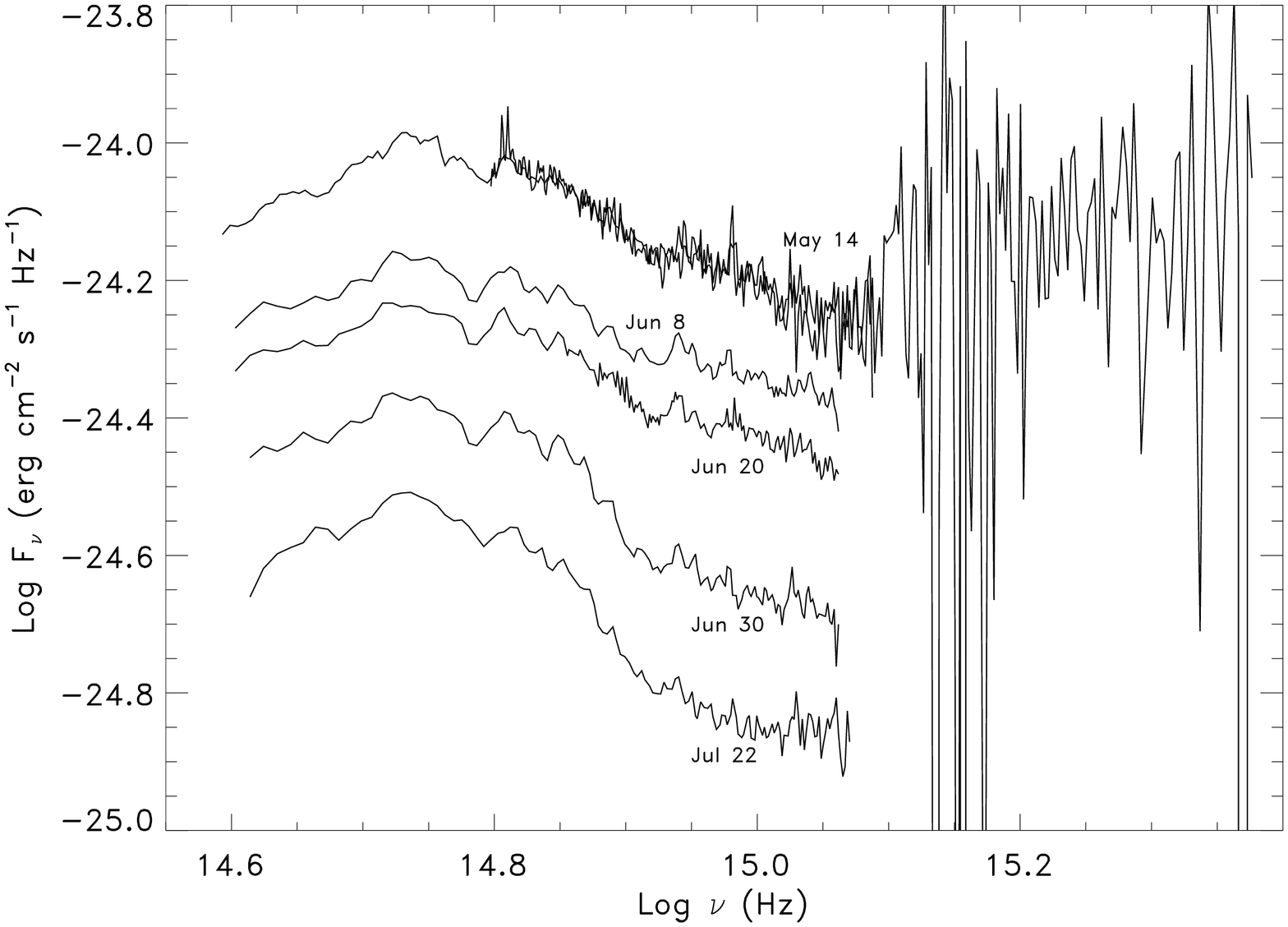}
\caption{HST optical/UV spectra dereddened assuming the Seaton (1979)
         extinction curve and $E_{B-V}=1.2$.  The far-UV spectrum is
         G160L data.  At lower frequencies, the spectra are composites
         of blue and red prism data, together with G270H/G400H spectra
         on May 14.}
\label{ContinuumFig}
\end{minipage}
\hspace{\fill}
\begin{minipage}[t]{3.0in}
\epsfig{width=2.1in,angle=90,file=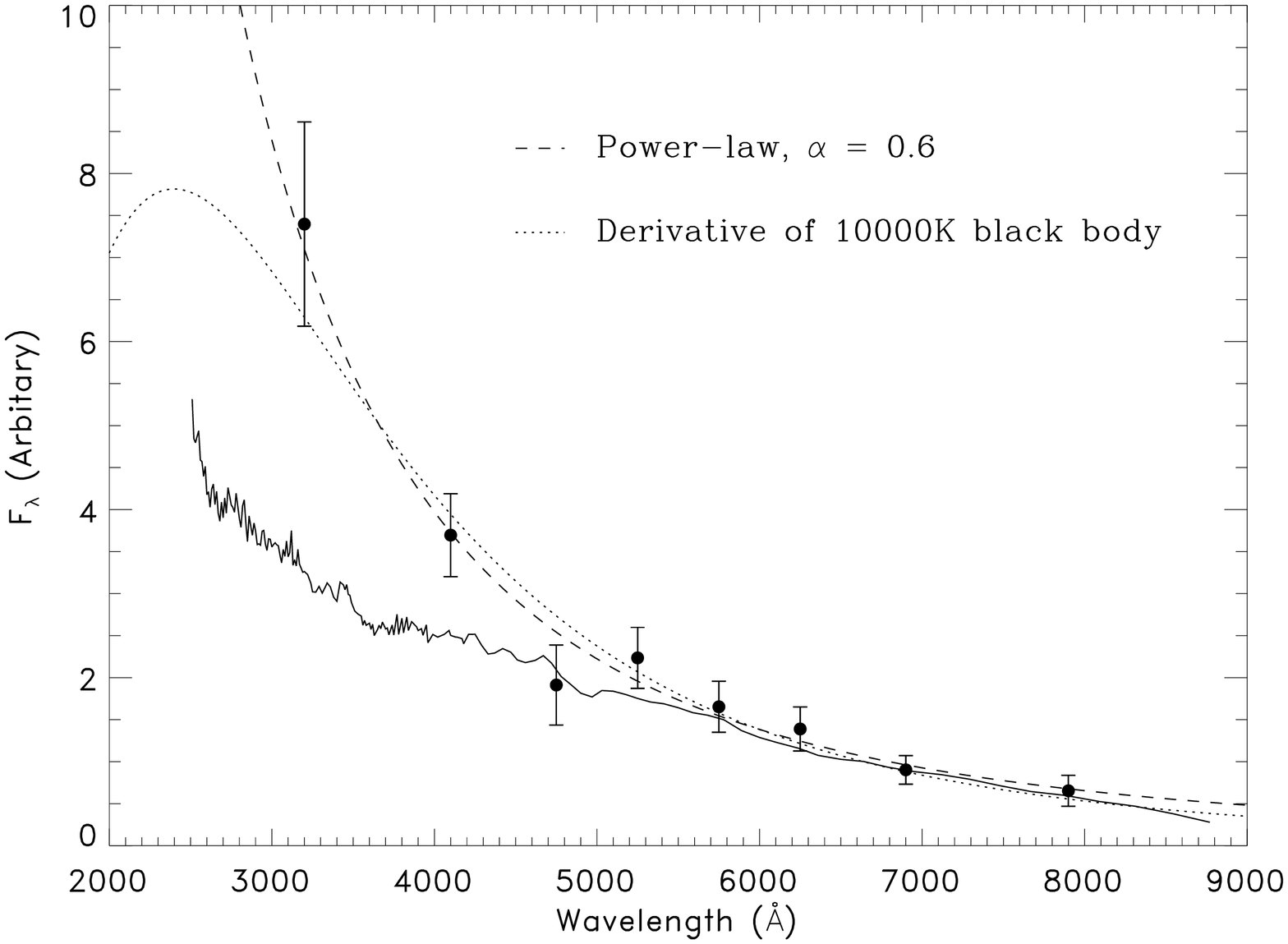}
\caption{A variability spectrum deduced from 1996 June 20 data, marked
         by points with error bars.  The solid line indicates the mean
         spectrum for comparison; the relative normalisation of the
         two is arbitrary.  Two models for the variability spectrum
         are also shown, as discussed in the text.}
\label{VarSpecFig}
\end{minipage}
\end{figure}
In Hynes et al.\ (1998b) we considered models for the optical/UV
continuum spectral evolution, shown in Fig.\ \ref{ContinuumFig}.  The
far-UV spectrum ($\log \nu > 15.1$) resembles the canonical
$\nu^{1/3}$ accretion disc spectrum.  We suggested two interpretations
for the spectrum at lower energies ($\log \nu < 15.1$): thermal
emission from the accretion disc, and non-thermal synchrotron emission
from a compact, self-absorbed source.  We will focus here on the
former model.  In this interpretation, we are seeing optically thick
thermal emission from an object with a temperature fixed around
9--10\,000\,K, shrinking in area.  Allowing for the contribution from
the bright F-type companion star, the area required is just about
consistent with a shrinking hot region of an accretion disc.  The
temperatures needed are consistent with gas just in the hot state of
the disc instability model (Cannizzo, Chen \& Livio 1995.)  The
shrinking area may thus be a signature of the long postulated cooling
wave of this model.

With this interpretation, there remains one important question to be
addressed: what is responsible for the heating of the disc?  The
spectrum that is seen is not purely the $\nu^{1/3}$ form expected for
an approximately steady-state, viscously heated accretion disc, and
instead is closer to the shape predicted for an irradiated disc
(Vrtilek et al. 1990), suggesting that it may be X-ray heating which is
keeping the disc in the hot state.
\section{RAPID SPECTROSCOPY -- What is the Spectrum Really Doing?}
So far, we have compressed our data in wavelength to produce light
curves and in time to examine spectra.  We will now seek a compromise
between spectral and temporal resolution, characterising the spectrum
of the variability observed.  This is only possible using very course
wavelength binning, as our signal-to-noise ratio is low, and hence
only a continuum variability spectrum can be estimated.  It is
immediately clear, however, that the variability spectrum is bluer
than the mean spectrum, i.e., there is a higher percentage variability
in the UV than at red wavelengths.  This variability spectrum can be
simply characterised as a blue power-law ($F_{\nu} \propto
\nu^{0.6}$); this is a similar slope to that of the power-law
component seen in the far-UV, and may indicate that this is the
variable component of the spectrum.  An alternative explanation,
however, is that a the blue variability spectrum is due to
reprocessing of X-ray variability.  The X-rays change the temperature
of the disc and hence we would expect the spectrum to be the {\it
derivative} of a black-body; the rate of change of the disc spectrum
with respect to temperature.  Our variability spectrum can be fit by
the derivative of a 10000\,K black-body, a comparable temperature to
that deduced from fitting the mean spectrum.  At present we cannot
discriminate between these possibilities, but we anticipate that
analysis of the remaining data segments will further constrain the
interpretation.
\section{Discussion -- Putting it all Together}
We have now accumulated several pieces of evidence which suggest
significant reprocessing of X-rays by the accretion disc: reprocessing
is seen directly through echo-mapping; the shape of the continuum
optical/UV spectrum is similar to that expected for an irradiated
disc; and the variability spectrum is consistent with the derivative
of the mean spectrum with respect to temperature.  It is natural to
ask whether the overall light curves could be reproduced by any
irradiation dominated model for the disc.

The most promising approach seems to be the following, hinging on the
X-ray spectral state evolution, from an initially soft, mainly thermal
spectrum towards the canonical two-component form (soft black-body
plus hard power-law).  The power-law component is widely believed to
be formed by Comptonisation of soft photons, so this change is
suggestive of an increasing optical depth of Comptonising material (a
similar scenario, but in reverse, was considered by Mineshige (1994)
for X-ray Nova Muscae).  If this material takes the form of a corona
above the surface of the disc, then it will affect X-ray irradiation
as follows.  Initially, soft X-rays will have a clear path to the
outer disc and irradiation will be efficient.  As the Comptonising
corona thickens, the Comptonised hard X-ray flux will increase, while
X-rays moving parallel to the disc will be increasingly attenuated,
shielding the outer disc and hence reducing the efficiency of
irradiation.  This decrease in efficiency dominates over the increase
in X-ray flux and so the reprocessed optical flux declines, even
though the X-rays are increasing.  The observed anti-correlation thus
emerges.  This model is not without difficulties; it requires the edge
of the disc to become partly shielded from X-rays, while our line of
sight to the X-ray source remains unobscured.  Since this is a high
inclination system, however, our line of sight does not lie very far
above the disc edge, and so some fine tuning of the shielding is
required.  We would also predict that irradiation should be strongest
on our first observation, May 14, and that we should thus see strong
echoes then.  In fact we see no optical/UV--X-ray correlations in the
May 14 data.

An alternative which therefore remains appealing is that the optical
energy budget is dominated by viscous heating, rather than by
irradiation; in this case, the reprocessing that is seen is merely a
perturbation to this, albeit a relatively large one, and does not
drive the long-term optical evolution.  Neither of these explanations
for the behaviour of GRO J1655--40 during 1996 that have been
suggested here can readily be rejected, and both have difficulties and
features to commend them.  Detailed modelling may be needed to rule
one, or even both, of them out.  There are clearly many questions
remaining to be answered concerning this outburst and this work
continues.
\section{Acknowledgements}
Many thanks to my collaborators on this work, co-authors on Hynes et
al.\ (1998a,b), and especially to my supervisor, Carole Haswell, for
endless constructive advice and encouragement.  Considerably less
progress would have been made on this project without the many
stimulating discussions I have had with other workers in the field too
numerous to list here.  I am supported by a PPARC Research
Studentship.  Support for this work was provided by NASA through grant
number GO-6017-01-94A from STScI, which is operated by AURA, under
NASA contract NAS5-26555 and also through contract NAS5-32490 for the
{\it RXTE} project.  This work made use of the {\it RXTE} and {\it
CGRO} Science Centers at NASA/GSFC and the NASA ADS Service.  Thanks
also to all at STScI for technical support.
\end{document}